# Polymorphism, crystal nucleation and growth in the phase-field crystal model in 2d and 3d

**Gyula I. Tóth,† György Tegze,† Tamás Pusztai,† and László Gránásy†‡§**

† Research Institute for Solid State Physics and Optics, P O Box 49, H−1525 Budapest, Hungary
‡ Brunel Centre for Advanced Solidification Technology, Brunel University, Uxbridge, UB8 3PH, UK
§ Corresponding author: grana@szfki.hu or Laszlo.Granasy@brunel.ac.uk

**Abstract.** We apply a simple dynamical density functional theory, the phase-field crystal (PFC) model of overdamped conservative dynamics, to address polymorphism, crystal nucleation, and crystal growth in the diffusion-controlled limit. We refine the phase diagram for 3d, and determine the line free energy in 2d, the height of the nucleation barrier in 2d and 3d for homogeneous and heterogeneous nucleation by solving the respective Euler-Lagrange (EL) equations. We demonstrate that in the PFC model, the body-centered cubic (bcc), the face-centered cubic (fcc), and the hexagonal close packed structures (hcp) compete, while the simple cubic structure is unstable, and that phase preference can be tuned by changing the model parameters: close to the critical point the bcc structure is stable, while far from the critical point the fcc prevails, with an hcp stability domain in between. We note that with increasing distance from the critical point the equilibrium shapes vary from sphere to the specific faceted shapes: rhombic-dodecahedron (bcc), truncated-octahedron (fcc), and hexagonal prism (hcp). Solving the equation of motion of the PFC model supplied with conserved noise, solidification starts with the nucleation of an amorphous precursor phase, into which the stable crystalline phase nucleates. The growth rate is found to be time dependent and anisotropic, which anisotropy depends on the driving force. We show that due to the diffusion-controlled growth mechanism, which is especially relevant for crystal aggregation in colloidal systems, dendritic growth structures evolve in large-scale isothermal single-component PFC simulations. Finally, we present results for eutectic solidification in a binary PFC model.

## 1. Introduction

Highly undercooled liquids often solidify to metastable (MS) crystal structures (Herlach 1994, Herlach *at al* 2007). The crystal structure is selected in the early nucleation stage of solidification, in which crystallike heterophase fluctuations form that drive the non-equilibrium liquid towards freezing. Heterophase fluctuations larger than a critical size, determined by the interplay of the interface free energy and the thermodynamic driving force, tend to grow, while the smaller ones decay with a high probability. Molecular dynamics (MD) simulations for the Lennard-Jones system show that various local structures such as icosahedral, face-centered cubic (fcc), hexagonal close packed (hcp) and body centered cubic (bcc) compete during solidification (Swope and Andersen 1990). Atomistic simulations imply that, in agreement with Ostwald's step rule, frequently that MS phase nucleates, whose structure lies the closest to the structure of the liquid (ten Wolde and Frenkel 1999). Indeed there are theoretical expectations that in simple liquids the first nucleating phase has the bcc structure (Alexander and McTague 1978, Klein 2001), an expectation supported by atomistic simulations for the Lennard-Jones system (ten Wolde *et al* 1995, 1996) and by experiments showing metastable bcc nucleation in supersaturated superfluid $^4$He, in preference to the stable hcp phase (Johnson and Elbaum 2000). Results from atomistic theory based on the density functional technique (DFT) suggest that crystallization might happen via a dense liquid / amorphous precursor phase (Lutsko and Nicolis 2006, Berry *et al* 2008a), a phenomenon reminiscent to the two-step transition seen in colloidal systems in 2d (Zhang and Liu 2007, Savage and Dinsmore 2009, DeYoreo 2010). In 3d colloidal systems crystallization to the random hexagonal close packed (rhcp) structure happens via a precursor of tiny compressed objects displaying only partial or embryonic crystal structure, missing long-range order (Schöpe *et al* 2006, 2007, Iacopini *et al* 2009a, 2009b). Other theoretical work implies that the presence of a metastable fluid critical point might assist crystal nucleation via a dense liquid precursor (ten Wolde and Frenkel 1997, Talanquer and Oxtoby 1998, Sear 2001, Shiryayev and Gunton 2004, Tóth and Gránásy 2007). These findings suggest that the two-step crystal nucleation via a precursor phase is a fairly general phenomenon both in 2d and 3d. The respective precursor phase may be amorphous or crystalline, depending on the multiplicity of metastabe phases available for the system. We note nevertheless that in other simple liquid such as the hard sphere liquid no sign of any precursor phase has been observed (Auer and Frenkel 2001a, 2001b, 2003).

    Heterogeneities such as container walls, floating solid particles, and free surfaces may assist the formation of the hetero-phase fluctuations: their presence may induce ordering in the liquid (Yasuoka *at al* 2000, Webb *et al* 2003, Auer and Frenkel 2003, Wang *et al* 2007). This ordering either helps or prevents the formation of heterophase fluctuations (Esztermann and Löwen 2005). When the ordering is compatible with the crystal structure to which the liquid freezes, the formation of heterophase fluctuations is enhanced at the wall, a phenomenon termed *heterogeneous nucleation*, as opposed to *homogeneous nucleation*, where the only heterogeneities in the liquid are its internal fluctuations. Heteroge-

Table 1. Classical nucleation theory for homogenous and heterogeneous processes in 2d and 3d.

| Dimensions | Shape | $W^*$ | Critical size | $f(\vartheta)$ |
|---|---|---|---|---|
| 2 | Circle | $\pi \cdot \gamma_{SL}^2 / \Delta\omega$ | $R^* = \gamma_{SL}/\Delta\omega$ | $[\vartheta - \tfrac{1}{2}\sin(2\vartheta)]/\pi$ |
|   | Hexagon | $2 \cdot 3^{1/2} \cdot \gamma_{SL}^2/\Delta\omega$ | $a^* = 2 \cdot \gamma_{SL}/(3^{1/2}\Delta\omega)$ | |
| 3 | Sphere | $(16\pi/3) \cdot \gamma_{SL}^3/\Delta\omega^2$ | $R^* = 2 \cdot \gamma_{SL}/\Delta\omega$ | $\tfrac{1}{4} \cdot [2 - 3\cos(\vartheta) + \cos(\vartheta)^3]$ |

Notation: $W^*$ — nucleation barrier, $R^*$ — critical radius, $a^*$ — critical edge length, $f$ — catalytic potency factor, $\vartheta$ — contact angle, $\gamma_{SL}$ — solid-liquid interface/line free energy, $\Delta\omega$ — thermodynamic driving force (grand potential density difference).

neous nucleation depends on such atomistic details as the structure of the wall, its chemical properties, surface roughness, and ordering of the liquid at the wall, etc. In the classical approach to heterogeneous nucleation these details are buried into the *equilibrium contact angle* $\vartheta$, which in turn reflects the relative magnitudes of the wall-solid ($\gamma_{WS}$), wall-liquid ($\gamma_{WL}$), and solid-liquid ($\gamma_{SL}$) interfacial free energies (e.g., Herring 1951): $\cos\vartheta = (\gamma_{WS} - \gamma_{WL})/\gamma_{SL}$. It relies on the droplet or capillarity approximation that neglects the anisotropy of the interfacial free energies, and regards the interfaces as mathematically sharp. Some predictions of the classical theory for 2d and 3d, we are going to refer to later, are compiled in table 1. While the classical model of heterogeneous nucleation captures some trends qualitatively (see e.g., Christian 1981), it is accurate for only large sizes where the thickness of the interface is indeed negligible relative to the size of the nucleus. In most cases, however, the size of nuclei is comparable to the interface thickness, casting doubts on the accuracy of the classical droplet model. Indeed in the case of homogeneous nucleation in the hard-sphere system, the droplet model fails under the conditions accessible for atomistic simulations (Auer and Frenkel 2001a). A practically important limit, in which quantitative predictions are possible for particle-induced crystallization is, when the particles are ideally wet by the crystalline phase, i.e., nucleation is avoided and the conditions of free growth limit the ability of a particle to start crystallization; a phenomenon studied extensively by Greer and co-workers (Geer *et al* 2000, Quested and Greer 2005, Reavley and Greer 2008).

Modeling of the interaction between the substrate and the solidifying liquid requires an atomistic approach. Molecular dynamics and Monte Carlo have provided important information on the microscopic aspects of the wetting of foreign walls by liquid and crystal (Toxwaerd 2002, Webb *et al* 2003, Auer and Frenkel 2003, Esztermann and Löwen 2005). Another atomistic technique, the dynamical density functional (DDFT) theory, has been used to address the effect of varying the structure of crystalline seeds on the process of crystallization (van Teeffelen *et al* 2008). Adaptation of a simple DDFT-type approach, the phase-field crystal (PFC) model (Elder *et al* 2002, Elder and Grant 2004), to elongated molecules has been used to study heterogeneous nucleation on unstructured walls (Prieler *et al* 2009). Pattern formation on periodic substrates represented by external potentials has also been studied by 2d PFC simulations (Achim *et al* 2006). Extension of such microscopic studies to other aspects of crystal nucleation (Gránásy *et al* 2010) is expected to create knowledge useful for establishing nucleation-controlled solidification and micro-patterning. Finally, it is also of considerable interest to see how far one can get with PFC type atomistic simulations when addressing complex larger scale growth forms including dendrites and eutectic structures.

Herein, we apply the PFC approach to investigate crystal nucleation and growth in 2d and 3d and to address (i) the phase diagram of the 3d PFC/Swift-Hohenberg model; (ii) the height of the nucleation for homogeneous and heterogeneous nucleation; (iii) equilibrium shapes for the 3d polymorphs; (iv) the existence of an amorphous precursor phase in homogeneous nucleation; (v) the effective interparticle potential the PFC model realizes; and (vi) the formation of dendritic and eutectic structures.

## 2. Phase-field crystal (PFC) models

The phase-field crystal model is a simple dynamical density functional theory of crystalline solidification developed by Elder and co-workers (2002). It represents the local state of matter by a time averaged particle density field, which is uniform in the liquid phase and periodic in the crystalline phase. It is based on a free energy functional that can be deduced (Elder and Grant 2004) from the perturbative density functional theory by Ramakrishnan and Yussouff (1979). After some simplifications one arrives to a Brazowskii/Swift-Hohenberg type free energy functional (Brazowskii 1975, Swift and Hohenberg 1977), while an overdamped conservative equation of motion is adopted to describe the time evolution of the particle density field. The relationship between the dynamical density functional theory and the PFC model has been addressed in detail by van Teeffelen *et al* (2009). In the past couple of years, the PFC model has been used successfully to address a broad range of phenomena such as elasticity and grain boundaries (Elder *et al* 2002), the anisotropy of the interfacial free energy (Wu and Karma 2007, Majaniemi and Provatas 2009) and growth rate (Tegze *et al* 2009b), dendritic and eutectic growth (Elder *et al* 2007, Provatas *et al* 2007, Pusztai *et al* 2008, Tegze *et al* 2009a), glass formation (Berry *et al* 2008a), melting at dislocations and grain boundaries (Berry *et al* 2008b, Mellenthin *et al* 2008), and polymorphism (Tegze *et al* 2009b). Although the PFC model is a microscopic approach, it has the advantage over other classical microscopic techniques, such as molecular dynamics simulations that the time evolution of the system can be studied on the many orders of magnitude longer diffusive time scale, making accessible the long-time behavior and the large-scale structures. It is worth emphasizing that the diffusion-controlled relaxation dynamics the PFC model assumes is indeed relevant for micron-scale colloidal systems (van Teeffelen *et al* 2008, 2009), where the self-diffusion of the particles is expected to be the dominant way of density relaxation. For normal liquids at small undercoolings the acoustic mode of density relaxation dominates, a phenomenon, which might be approximately incorporated into the PFC model by adding a term proportional to $\partial^2 n/\partial t^2$ (Majaniemi 2009).

*2.1. The single-component phase-field crystal model*

*2.1.1. The free energy functional.* The free energy of the PFC model can be derived (see Elder and Grant 2004) from the perturbative density functional theory of Ramakrishnan and Yussouff (1979), in which the free energy difference $\Delta F = F - F_L^{ref}$ of the crystal relative to a reference liquid (of particle density $\rho_L^{ref}$) is expanded with respect to the density difference $\Delta \rho = \rho - \rho_L^{ref}$ between the crystal and the liquid, retaining the terms up to the two-particle term:

$$\frac{\Delta F}{kT} = \int d\mathbf{r} \left[ \rho \ln\left(\frac{\rho}{\rho_L^{ref}}\right) - \Delta \rho \right] - \frac{1}{2} \iint d\mathbf{r}_1 d\mathbf{r}_2 [\Delta \rho(\mathbf{r}_1) C(\mathbf{r}_1, \mathbf{r}_2) \Delta \rho(\mathbf{r}_2)] + \ldots \quad (1)$$

where $C(\mathbf{r}_1, \mathbf{r}_2) = C(|\mathbf{r}_1 - \mathbf{r}_2|)$ is the two-particle direct correlation function of the reference liquid. Writing the particle density a Fourier expanded form, one obtains for the solid $\rho_S = \rho_L^{ref} \{1 + \eta_S + \sum_\mathbf{K} A_\mathbf{K} \cdot \exp(i\mathbf{Kr})\}$, where $\eta_S$ is the fractional density change upon freezing, while $\mathbf{K}$ are reciprocal lattice vectors, and $A_\mathbf{K}$ are the respective Fourier amplitudes. Introducing the reduced number density relative to the reference liquid, $n = (\rho - \rho_L^{ref})/\rho_L^{ref} = \eta_S + \sum_\mathbf{K} A_\mathbf{K} \cdot \exp(i\mathbf{Kr})$ one finds

$$\frac{\Delta F}{kT} = \int d\mathbf{r} \left[ \rho_L^{ref}(1+n)\ln(1+n) - \rho_L^{ref} n \right] - \frac{1}{2} \iint d\mathbf{r}_1 d\mathbf{r}_2 [\rho_L^{ref} n(\mathbf{r}_1) C(|\mathbf{r}_1 - \mathbf{r}_2|) \rho_L^{ref} n(\mathbf{r}_2)] + \ldots \quad (2)$$

Next, we expand $C(|\mathbf{r}_1 - \mathbf{r}_2|)$ in Fourier space, $\hat{C}(k) \approx \hat{C}_0 + \hat{C}_2 k^2 + \hat{C}_4 k^4 + \ldots$. Note that $\hat{C}(k)$ has its 1st peak at $k = 2\pi/\sigma$, and the sign of the coefficients is expected to alternate, while $\sigma$ is the inter-particle distance. Defining the dimensionless form of $\hat{C}(k)$ as $c(k) = \rho_L^{ref} \hat{C}(k) \approx \sum_{j=0}^{m} c_{2j} k^{2j} = \sum_{j=0}^{m} b_{2j} (k\sigma)^{2j}$, which is thus related to the structure factor as $S(k) = 1/[1 - c(k)]$. Considering these, integrating the second term on RHS with respect to $\mathbf{r}_2$ and replacing $\mathbf{r}_1$ by $\mathbf{r}$, the free energy difference reads as

$$\frac{\Delta F}{kT \rho_L^{ref}} \approx \int d\mathbf{r} \left[ (1+n)\ln(1+n) - n - \frac{n}{2} \left\{ \sum_{j=0}^{m} (-1)^j c_{2j} \nabla^{2j} \right\} n \right]. \quad (4)$$

The reference liquid (of particle density $\rho_L^{ref}$) is not necessarily the initial liquid. Thus, we may have here two parameters to control the driving force for solidification: the initial liquid number density $n_L^0$, and the temperature, if the direct correlation function depends on temperature. Taylor-expanding $\ln(1 + n)$ for small $n$ one obtains

$$\frac{\Delta F}{kT \rho_L^{ref}} \approx \int d\mathbf{r} \left[ \frac{n^2}{2} - \frac{n^3}{6} + \frac{n^4}{12} - \frac{n}{2} \left\{ \sum_{j=0}^{m} (-1)^j c_{2j} \nabla^{2j} \right\} n \right]. \quad (5)$$

For $m = 2$, corresponding to the simplest version of PFC (Elder *et al* 2002), and taking the alternating sign of the expansion coefficients of $\hat{C}_i$ into account, equation (5) transforms to the following form:

$$\Delta F \approx kT \rho_L^{ref} \int d\mathbf{r} \left\{ \frac{n^2}{2}(1+|b_0|) + \frac{n}{2}[|b_2|\sigma^2 \nabla^2 + |b_4|\sigma^4 \nabla^4]n - \frac{n^3}{6} + \frac{n^4}{12} \right\}. \quad (6)$$

Introducing the new variables

$B_L = 1 + |b_0| = 1 - c_0$      [$= (1/\kappa)/(\rho_L^{ref} kT)$, where $\kappa$ is the compressibility],
$B_S = |b_2|^2/(4|b_4|)$      [$= K/(\rho_L^{ref} kT)$, where $K$ is the bulk modulus of the crystal],
$R = \sigma(2|b_4|/|b_2|)^{1/2}$      [$=$ the new length scale ($x = R \cdot \tilde{x}$), which is now related to the position of the maximum of the Taylor expanded $\hat{C}(k)$], and a multiplier $v$ for the $n^3$ term (that accounts for the 0th order contribution from 3-particle correlation), one obtains the form used by Berry *at al* (2008a, 2008b):

$$\Delta F = \int d\mathbf{r} I(n) = kT \rho_L^{ref} \int d\mathbf{r} \left\{ \frac{n}{2}[B_L + B_S(2R^2\nabla^2 + R^4\nabla^4)]n - v\frac{n^3}{6} + \frac{n^4}{12} \right\}, \quad (7)$$

where $I$ stands for the full (dimensional) free energy density.

*The Swift-Hohenberg type dimensionless form.* It can be shown that introducing the set of new variables $x = R \cdot \tilde{x}$, $n = (3B_S)^{1/2} \psi$, $\Delta F = (3 \rho_L^{ref} kTR^d B_S^2) \cdot \Delta \tilde{F}$, the free energy functional transcribes into a Swift-Hohenberg form:

$$\Delta \tilde{F} = \int d\tilde{\mathbf{r}} \left\{ \frac{\psi}{2} \left[ r^* + (1 + \tilde{\nabla}^2)^2 \right] \psi + t^* \frac{\psi^3}{3} + \frac{\psi^4}{4} \right\}, \tag{8}$$

where, $t^* = -(v/2) \cdot (3/B_S)^{1/2} = -v \cdot (3|b_4|/|b_2|^2)^{1/2}$ and $r^* = \Delta B/B_S = (1 + |b_0|)/[|b_2|^2/(4|b_4|)] - 1$, while $\psi = n/(3B_S)^{1/2}$. The quantities involved in equation (8) are all dimensionless. The form of the free energy suggests that the $m = 2$ PFC model contains two dimensionless similarity parameters $r^*$ and $t^*$ composed of the original model parameters. Remarkably, even the third-order term can be eliminated. In the respective $t^{*'} = 0$ Swift-Hohenberg model, the state $[r^{*'} = r^* - (t^*)^2/3, \psi' = \psi - t^*/3]$ corresponds to the state $(r^*, \psi)$ of the original $t^* \neq 0$ model. This transformation leaves the grand canonical potential difference, the Euler-Lagrange equation and the equation of motion invariant. Accordingly, it is sufficient to address the $t^* = 0$ case, as we do in the rest of this work.

*Eight-order fitting of C(k) (PFC EOF):* Jaatinen *et al.* (2009) have recently proposed an eight-order expansion of the Fourier transform of the direct correlation function around its maximum ($k = k_m$):

$$C(k) \approx C(k_m) - \Gamma \left( \frac{k_m^2 - k^2}{k_m^2} \right)^2 - E_B \left( \frac{k_m^2 - k^2}{k_m^2} \right)^4. \tag{9}$$

The expansion parameters were then fixed so that the liquid compressibility and the position, height, and the second derivative of $C(k)$ are accurately recovered. This is ensured by

$$\Gamma = -\frac{k_m^2 C''(k_m)}{8} \quad \text{and} \quad E_B = C(k_m) - C(0) - \Gamma. \tag{10}$$

With this choice of the model parameters and relevant data for Fe by (Wu and Karma 2007) they reported a fair agreement with MD results for the volume change upon melting, the bulk moduli of the liquid and solid phases, and the magnitude and anisotropy of the solid-liquid interfacial free energy (Jaatinen *et al* 2009).

*2.1.2. The equation of motion.* Similarly to the DDFT for colloidal systems (van Teeffelen *et al* 2008, 2009), an overdamped conserved dynamics is assumed here, however, with a constant mobility coefficient of $M_\rho = \rho_0 D_\rho / kT$. Accordingly, the (dimensional) equation of motion has the form

$$\frac{\partial \rho}{\partial t} = \nabla \left\{ M_\rho \left[ \nabla \frac{\delta \Delta F}{\delta \rho} \right] + \left( \frac{2kTM_\rho}{\Delta x^d \Delta t} \right)^{1/2} \mathscr{N} \right\}, \tag{11}$$

where the second term is the discretized form of the conserved noise (Karma and Rappel 1999) that applies here, while $\mathscr{N}$ is a Gaussian white noise of unit standard deviation. Changing from variable $\rho$ to $n$, introducing $M_n = [(1+n_0) D_\rho /(kT\rho_L^{ref})]$, and scaling the time and distance as $t = \tau \cdot \tilde{t}$ and $x = \sigma \cdot \tilde{x}$, where $\tau = \sigma^2/[D_\rho (1 + n_0)]$, and inserting the free energy from equation (6), one obtains the following dimensionless equation of motion:

$$\frac{\partial n}{\partial \tilde{t}} = \tilde{\nabla}^2 \left[ n(1 + |b_0|) + \sum_{j=1}^{m} |b_{2j}| \tilde{\nabla}^{2j} n - \frac{n^2}{2} + \frac{n^3}{3} \right] + \tilde{\nabla} \left( \frac{2}{\rho_L^{ref} \sigma^d \Delta \tilde{x}^d \Delta \tilde{t}} \right)^{1/2} \mathscr{N}. \tag{12}$$

Analogously, the equation of motion corresponding to equation (7), has the form.

$$\frac{\partial n}{\partial t} = \nabla \left\{ M_n \nabla \left[ (kT\rho_L^{ref}) \left( [B_L + B_S(2R^2 \nabla^2 + R^4 \nabla^4)] n - v \frac{n^2}{2} + \frac{n^3}{3} \right) \right] + \left( \frac{2kTM_n}{\Delta x^d \Delta t} \right)^{1/2} \mathscr{N} \right\}. \tag{13}$$

*The Swift-Hohenberg type dimensionless form:* Introducing the set of new variables $t = \tau \cdot \tilde{t}$, $x = R \cdot \tilde{x}$, and $n = (3B_S)^{1/2} \psi = (3B_S)^{1/2} [\psi' + t^*/3]$ into equation (13), where $\tau = R^2/(B_S M_n \rho_L^{ref} kT)$, the equation of motion can be written in the form (Elder *et al* 2002, Elder and Grant 2004)

$$\frac{\partial \psi'}{\partial \tilde{t}} = \tilde{\nabla}^2 \left\{ \left[ r^{*'} + (1 + \tilde{\nabla}^2)^2 \right] \psi' + \psi'^3 \right\} + \tilde{\nabla} \left( \frac{\alpha^*}{\Delta \tilde{x}^d \Delta \tilde{t}} \right)^{1/2} \mathscr{N}, \tag{14}$$

where $r^{*'} = r^* - (t^*)^2/3 = [\Delta B - (v/2)^2]/B_S = (1 + |b_0|)/[|b_2|^2/(4|b_4|)] - [1 + v^2 \cdot (|b_4|/|b_2|^2)]$ and the dimensionless noise strength is $\alpha^* = 2/(3B_S^2 \rho_L^{ref} R^d) = 2^{5-d/2} |b_4|^{2-d/2}/[3\sigma^d \rho_L^{ref} |b_2|^{4-d/2}]$, while the correlator for the dimensionless noise reads as $\langle \zeta(\tilde{\mathbf{r}},\tilde{t}), \zeta(\tilde{\mathbf{r}}',\tilde{t}') \rangle = \alpha^* \cdot \tilde{\nabla}^2 \delta(\tilde{\mathbf{r}} - \tilde{\mathbf{r}}') \cdot \delta(\tilde{t} - \tilde{t}')$.

Summarizing, the dynamical $m = 2$ PFC model has two dimensionless similarity parameters $r^{*\prime}$ and $\alpha^*$ composed of the original (physical) model parameters. This is the generic form of the $m = 2$ PFC model; some other formulations (Elder and Grant 2004, Berry et al 2008a, 2008b) can be transformed into this form.

Simulation of nucleation using the equation of motion is non-trivial due to several effects (see e.g. Haataja et al 2010, Plapp 2010). In the DDFT type models, nucleation does not occur from a homogeneous initial fluid state unless adding Langevin noise to the equation of motion to represent the thermal fluctuations. This is, however, not without conceptual difficulties, as pointed out in a discussion by several authors (Marconi and Tarazona 1999, Löwen 2003, Archer and Rauscher 2004): viewing the number density a quantity that has been averaged over the ensemble, all the fluctuations are (in principle) incorporated into the free energy functional; via adding noise to the equation of motion part of the fluctuations is counted doubly (Marconi and Tarazona 1999, Löwen 2003). If, on the other hand, the number density is assumed to be coarse-grained in time, there is phenomenological motivation to add the noise to the equation of motion (Archer and Rauscher 2004). The latter approach is appealing in several ways: crystal nucleation is feasible from a homogeneous state and capillary waves appear at the crystal-liquid interface. Since in the present study our aim is to investigate how nucleation and growth happen on the atomistic level, we incorporate a conserved noise term into the equation of motion (see equations (11)–(14)). To overcome some difficulties occurring when discretizing the noise (Plapp 2010), we use here colored noise obtained by filtering out the unphysical short wavelengths smaller than the interparticle distance (this removes both the ultraviolet catastrophe, expected in 3d (Karma 2009), and the associated dependence of the results on spatial resolution).

*2.1.3. The Euler-Lagrange equation.* The EL equation has the form

$$\frac{\delta \Delta \tilde{F}}{\delta \psi} = \frac{\delta \Delta \tilde{F}}{\delta \psi}\bigg|_{\psi_0} . \quad (15)$$

Here $\psi_0$ is the reduced particle density of the reference liquid, while a no-flux boundary condition is prescribed at the borders of the simulation window ($\mathbf{n}\nabla\psi = 0$ and $(\mathbf{n}\cdot\nabla)\Delta\psi = 0$, where $\mathbf{n}$ is the normal vector of the boundary). Inserting the free energy functional, and rearranging the terms, one arrives to

$$[r^* + (1+\nabla^2)^2](\psi - \psi_0) = t^*(\psi^2 - \psi_0^2) - (\psi^3 - \psi_0^3) . \quad (16)$$

Equation (16) together with the boundary condition represents a 4$^{th}$ order boundary value problem (BVP).

*2.1.4. Modeling of a crystalline substrate in 2d.* In the region filled by the substrate, we add an external potential term $V\psi$ to the free energy density. We chose the following forms for the potential in 2d and 3d, respectively: $V(x, y) = V_0 + V_1[\cos(qx) + \cos(qy)]$, where $q = 2\pi/a_0$ and $a_0$ is the lattice constant of the external potential. When these potentials are strong enough, these potentials can enforce the particles to realize the otherwise unstable square-lattice structure.

*2.2. The binary phase-field crystal model*

*2.2.1. The free energy functional.* In derivation of the binary PFC model, the starting point is the free energy functional of the binary perturbative density functional theory, where the free energy is Taylor expanded relative to the liquid state (denoted by subscript L) up to 2nd order in density difference (up to two-particle correlations) (Elder et al 2007):

$$\frac{\Delta F}{kT} = \int d\mathbf{r}\left[\rho_A \ln\left(\frac{\rho_A}{\rho_A^L}\right) - \Delta\rho_A + \rho_B \ln\left(\frac{\rho_B}{\rho_B^L}\right) - \Delta\rho_B\right] \\ -\frac{1}{2}\iint d\mathbf{r}_1 d\mathbf{r}_2 [\Delta\rho_A(\mathbf{r}_1)C_{AA}(\mathbf{r}_1,\mathbf{r}_2)\Delta\rho_A(\mathbf{r}_2) + \Delta\rho_B(\mathbf{r}_1)C_{BB}(\mathbf{r}_1,\mathbf{r}_2)\Delta\rho_B(\mathbf{r}_2) + 2\Delta\rho_A(\mathbf{r}_1)C_{AB}(\mathbf{r}_1,\mathbf{r}_2)\Delta\rho_B(\mathbf{r}_2)] \quad (17)$$

where $k$ is Boltzmann's constant, $\Delta\rho_A = \rho_A - \rho_A^L$ and $\Delta\rho_B = \rho_B - \rho_B^L$. It is assumed here that all two point correlation functions are isotropic, i.e., $C_{ij}(\mathbf{r}_1, \mathbf{r}_2) = C_{ij}(|\mathbf{r}_1 - \mathbf{r}_2|)$. Taylor expanding direct correlation functions in Fourier space up to 4th order, one obtains $C_{ij} = [C^0_{ij} - C^2_{ij}\nabla^2 + C^4_{ij}\nabla^4]\delta(\mathbf{r}_1 - \mathbf{r}_2)$ in real space, where $\nabla$ differentiates with respect to $\mathbf{r}_2$ (see Elder et al 2007). The partial direct correlation functions $C_{ij}$ can be related to measured or computed partial structure factors (see e.g. Woodhead-Galloway and Gaskell 1968).

Following Elder et al (2007), we introduce the reduced partial particle density differences $n_A = (\rho_A - \rho_A^L)/\rho_L$ and $n_A = (\rho_B - \rho_B^L)/\rho_L$, where $\rho_L = \rho_A^L + \rho_B^L$. It is also convenient to introduce the new variables $n = n_A + n_B$ and $(\delta N) = (n_B - n_A) + (\rho_B^L - \rho_A^L)/\rho_L$. Then, expanding the free energy around $(\delta N) = 0$ and $n = 0$ one obtains

$$\frac{\Delta F}{\rho_L kT} = \int d\mathbf{r}\left\{\frac{n}{2}[B_L + B_S(2R^2\nabla^2 + R^4\nabla^4)]n + \frac{t}{3}n^3 + \frac{v}{4}n^4 + \gamma(\delta N) + \frac{w}{2}(\delta N)^2 + \frac{u}{4}(\delta N)^4 + \frac{L^2}{2}|\nabla(\delta N)|^2 + ...\right\} . \quad (18)$$

*2.2.2. The equations of motion.* It is assumed that the same $M$ mobility applies for the two species A and B (corresponding to substitutional diffusion) that decouples the dynamics of $n$ and ($\delta N$) fields. Assuming, furthermore, a constant $M_e$ mobility and conserved dynamics, the equations of motions for the two fields have the form (Elder *et al* 2007):

$$\frac{\partial n}{\partial t} = M_e \nabla^2 \frac{\delta \Delta F}{\delta n} \qquad \text{and} \qquad \frac{\partial (\delta N)}{\partial t} = M_e \nabla^2 \frac{\delta F}{\delta (\delta N)}, \qquad (19)$$

while the respective effective mobility is $M_e = 2M/\rho^2$. Taylor expanding then $B_L$, $B_S$ and $R$ in terms of ($\delta N$) of coefficients $B_j^L$, $B_j^S$ and $R_j$, retaining only coefficients $B_0^L$, $B_2^L$, $B_0^S$, $R_0$ and $R_1$, and inserting the free energy (equation 18) into equations (19), one obtains

$$\frac{\partial n}{\partial t} = M_e \nabla^2 \left[ \begin{array}{l} n\{B_0^L + B_2^L(\delta N)^2\} + tn^2 + vn^3 + \frac{B_0^S}{2}\{2[R_0 + R_1(\delta N)]^2 \nabla^2 + [R_0 + R_1(\delta N)]^4 \nabla^4\}n \\ + \frac{B_0^S}{2}\{2\nabla^2(n[R_0 + R_1(\delta N)]^2) + \nabla^4(n[R_0 + R_1(\delta N)]^4)\} \end{array} \right], \qquad (20a)$$

$$\frac{\partial (\delta N)}{\partial t} = M_e \nabla^2 \left[ \begin{array}{l} B_2^L(\delta N)n^2 + 2B_0^S n\{[R_0 + R_1(\delta N)]R_1 \nabla^2 + [R_0 + R_1(\delta N)]^3 R_1 \nabla^4\}n \\ + \gamma + w(\delta N) + u(\delta N)^3 - L^2 \nabla^2 (\delta N) \end{array} \right]. \qquad (20b)$$

*2.2.3. The Euler-Lagrange equations.* The extremum of the grand potential functional requires that its first functional derivatives are zero, i.e.

$$\frac{\delta \Delta F}{\delta n} = \frac{\delta \Delta F}{\delta n}\bigg|_{n_0, \delta N_0} \qquad \text{and} \qquad \frac{\delta F}{\delta (\delta N)} = \frac{\delta F}{\delta (\delta N)}\bigg|_{n_0, \delta N_0}, \qquad (21)$$

where $n_0$ and $\delta N_0$ are the total and differential particle densities for the (homogeneous) reference state. Inserting equations (20a) and (20b) into equation (21), after rearranging one obtains

$$\left[ B_L(\delta N) + B_S R(\delta N)^2 \{2\nabla^2 + R(\delta N)^2 \nabla^4\} + \frac{B_S}{2}\{\nabla^2[2R^2] + \nabla^4[R^4]\} \right](n - n_0) = -t(n^2 - n_0^2) - v(n^3 - n_0^3) \qquad (22a)$$

$$L^2 \nabla^2 (\delta N) = \frac{\partial B_L}{\partial (\delta N)}[(\delta N)n^2 - (\delta N)_0 n_0^2] + w[(\delta N) - (\delta N)_0] + u[(\delta N)^3 - (\delta N)_0^3] + 2B_S R \frac{\partial R}{\partial (\delta N)} n(\nabla^2 + R^2 \nabla^2)n \qquad (22b)$$

These equations are to be solved assuming no-flux boundary conditions at the border of the simulation box for both fields ($\mathbf{n}\nabla n = 0$, $(\mathbf{n}\cdot\nabla)\Delta n = 0$, $\mathbf{n}\nabla(\delta N) = 0$ and $(\mathbf{n}\cdot\nabla)\Delta(\delta N) = 0$).

*2.3. Solution of the equations of motion and the Euler-Lagrange equations.*

These equations of motion have been solved numerically on uniform rectangular 2d and 3d grids using a fully spectral semi-implicit scheme described in (Tegze *et al* 2009a) and periodic boundary condition at the perimeters. A parallel C code relying on the MPI protocol has been developed. To optimize the performance, we have developed a parallel FFT code based on the FFTW3 library (Frigo and Johnson 2005).

The EL equations have been solved here numerically, using a semi-spectral successive approximation scheme combined with the operator-splitting method (Tóth and Tegze 2010).

The numerical simulations presented in this paper have been performed on three computer clusters: One, that consists of 24 nodes, each equipped with two 2.33 GHz Intel processors of 4 CPU cores (192 CPU cores in all on the 24 nodes), 8 GB memory/node, and with 10 Gbit/s (InfiniBand) inter-node communication; a similar one with 16 nodes (128 CPU cores), and a third cluster, which consists of 36 similar nodes (288 CPU cores) with 24 GB memory/node, however, with 40 Gbit/s (InfiniBand) communication in between. The EL equations have been solved on three superservers, each consisting of 4 NVidia Tesla GPU cards with 4 GB memories/card and 48 GB system memory.

*2.5. Model parameters used.*

In 2d the computations have been performed at the reduced temperature $r^* = -0.5$, while $t^* = 0$. The corresponding coexisting densities obtained with full free energy minimization using the EL equation technique for the liquid and 2d hexagonal lattices are $\psi_L^e = -0.51398$ and $\psi_{Hex}^e = -0.38475$, respectively. This value of $r^*$ leads to a strongly faceted equilibrium shape and growth forms with excluded orientations (Gránásy *et al* 2010) closely resembling to those observed in 2d colloidal experiments (Onoda 1985, Skjeltorp 1987).

Unless stated otherwise, the 3d colloidal computations have been performed using a parameter set that has been chosen in a recent study so as to mimic characteristic features of charged colloidal systems (Tegze *et al* 2009b): $B_S =$

$3^{-1/2}/2$, $\Delta B = B_L − B_S = 5\times 10^{−5}$, and $v = 3^{1/4}/2$. Remarkably, with this choice of parameters the free energies of the bcc, fcc and hcp phases are very close to each other (Tegze *et al* 2009b) and the common tangent construction to the Helmholtz free energy density curves yielded the following liquid-solid coexistence regions: liquid–bcc: $−0.0862 < n_0 < −0.0315$, liquid–hcp: $−0.0865 < n_0 < −0.0344$, and liquid–fcc: $−0.0862 < n_0 < −0.0347$.

In the eight-order fitting PFC simulations for Fe, we have used the model parameters by Jaatinen *et al* (2009) referring to the melting point, however, we have increased the density to drive the liquid phase out of equilibrium.

In the binary simulations for 2d eutectic patterns the parameter set by Tegze *et al* (2009a) has been used, while in the 3d eutectic computations, the following parameters have been applied $B_0^L = 1.03$, $B_2^L = −1.8$, $B_0^S = 1$, $R_0 = 1$, $R_1 = 0$, $t = −0.6$, $v = 1$, $\gamma = 0$, $u = 4$, $w = −0.12$, and $L^2 = 4$.

## 3. Results and discussion

*3.1. Equilibrium features*

In this subsection we refine some sections of the phase diagram, and compute the equilibrium interfacial properties, the equilibrium shapes, and various properties of nuclei by solving the Euler-Lagrange equations numerically. Since in equilibrium the single-component PFC model is mathematically equivalent to the Swift-Hohenberg (SH) theory, the result presented in this section are equally valid for the latter.

In all cases, the numerical solution procedure has been started with an initial guess based on the single-mode approximation. For the bcc, fcc, and sc phases the respective normalized number densities were as follows: bcc: $\psi = 4A\{\cos(qx)\cdot\cos(qy) + \cos(qy)\cdot\cos(qz) + \cos(qx)\cdot\cos(qz)\}$ see Wu and Karma (2007); fcc: $\psi = 8A\{\cos(qx)\cdot \cos(qy)\cdot\cos(qz)\}$, and sc: $\psi = 2A\{\cos(qx) + \cos(qy) + \cos(qz)\}$, while the following *ansatz* by Gránásy and Tóth (Tegze *et al* 2009b) has been used for the hcp structure: $\psi = A\{\cos(2qy/\sqrt{3}) + \cos(qx − qy/\sqrt{3}) − \cos(2\pi/3 − qx + qy/\sqrt{3}) + \cos(qx + qy/\sqrt{3}) − \cos(−4\pi/3 + qx + qy/\sqrt{3}) − \cos(−2\pi/3 + 2qy/\sqrt{3})\}\cdot\cos\{(\sqrt{3}/\sqrt{8})qz\}$. Here $q = 2\pi/a$, while the lattice constant $a$ and the amplitude $A$ have been determined by analytic minimization of the free energy.

*3.1.1. Phase diagrams for the PFC/SH model (from EL equation).* While in the single component case, the 1d and 2d phase diagrams are fairly well known (Elder *et al* 2002, Elder and Grant 2004), and different versions of the 3d phase diagram have been presented by single-mode computations (Wu and Karma 2007) and by full free energy minimization (Jaatinen and Ala-Nissila 2010), we have reexamined the 3d phase diagram using the Euler-Lagrange technique: A single-mode initial guess has been applied for the scaled number density $\psi$ in a single cell of the crystal structure, which then has been solved by our numerical solver. A refined 3d phase diagram for the single-component case is shown in figure 1. It is in a general agreement with the results Jaatinen and Ala-Nissila (2010) obtained previously with a different method. It consists of a single domain for each of the bcc, hcp and fcc phases, where the given phase is stable. The three-phase equilibria (liquid-hcp-bcc, liquid-fcc-hcp, hcp-bcc-rod, and fcc-hcp-rod) are represented by horizontal peritectic lines in the phase diagram. Linear stability analysis of the liquid phase yields an instability region whose border is denoted by the heavy gray line in figure 1. The PFC/SH model predicts a critical point between the liquid and solid phases at ($r^* = 0$ and $\psi_0 = 0$). It is appropriate to mention in this respect that there is no convincing theoretical or experimental evidence for the existence of a critical point between crystalline and liquid phases in *simple* single component systems (Skripov 1976, Bartell and Wu 2007). Remarkably, however, a recent molecular dynamics study relying on a pair potential akin to the Derjaguin-Landau-Verwey-Overbeek (DLVO) potential with a secondary minimum (often used for modeling charged colloids) indicates the presence of a critical point between the solid and liquid phases (Elenius and Dzugutov 2009). We note finally that under the conditions, we use in our simulations, the driving force (the grand potential density difference $\Delta\omega_X = f_X(n_X) − \partial f_L/\partial n(n_0)\cdot[n_X − n_0] − f_L(n_0) = −\Delta p$ relative to the initial liquid, where $n_X$ is the crystal density that maximizes the driving force, and $\Delta p$ is the pressure difference relative to the liquid) is comparable for the bcc, fcc and hcp phases, though bcc is slightly preferred with the exception of a small region near the equilibrium liquid density, where the hcp phase has the largest driving force (Tegze *et al* 2009b). For larger densities, the hcp and fcc phases are metastable.

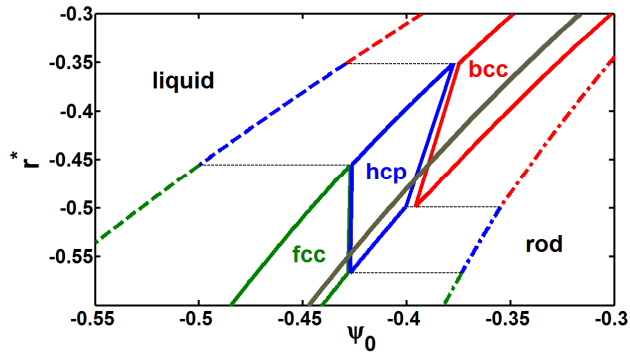

**Figure 1.** Solid-liquid coexistence in the phase diagram of the 3d PFC/SH model. The coexistence lines have been computed via solving the Euler-Lagrange equation. The liquid phase is unstable right of the heavy gray line.

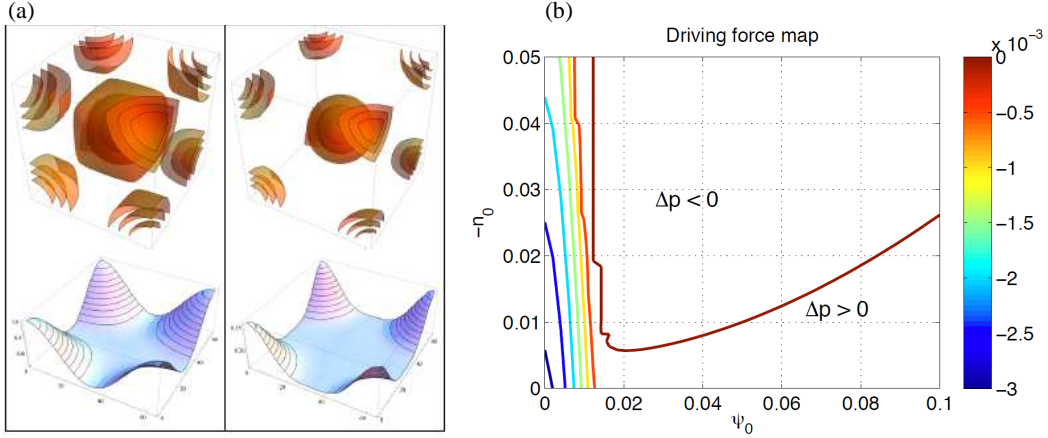

**Figure 2.** Thermodynamics of the 3d eutectic system in the two component PFC model: (a) Spatial distribution of the total ($n$, left) and differential ($\delta N$, right) reduced number densities after full free energy minimization performed using the Euler-Lagrange equation visualized with iso-surfaces (top) and by the in-plane distribution of the fields. (b) Contour map of the thermodynamic driving force for eutectic solidification as a function of the properties of the homogeneous initial liquid.

In the case of the binary system, we have used the EL equations to map the thermodynamic driving force

$$-\Delta p = \frac{\Delta F[n(\mathbf{r}), \delta N(\mathbf{r})]}{V} - f_0 - \left.\frac{\partial I}{\partial n}\right|_{(n_0, \delta N_0)} (\bar{n} - n_0) - \left.\frac{\partial I}{\partial (\delta N)}\right|_{(n_0, \delta N_0)} [\overline{\delta N} - \delta N_0] \tag{23}$$

as a function of the initial total reduced particle density ($n_0$) and the differential reduced particle density ($\delta N_0$). Here bars over the quantities denote averaging over the cell, while $I$ is the integrand of the Helmholtz free energy functional. The initial guess for the single-cell solution have been taken from the single-mode approximation for $n$, while a homogeneous initial $\delta N$ has been assumed. The converged fields are shown in figure 2a, while the driving force map is displayed in figure 2b. Note the narrow region, where eutectic solidification is preferable. Indeed, we have seen coupled eutectic solidification, when solving the equation of motion in this region.

*3.1.2. Equilibrium line free energy in the 2d PFC/SH model by solving the EL equation.* The solution of the EL equation has been obtained for the flat interface by starting from an initial guess of a liquid-solid-liquid sandwich of the equilibrium densities and a *tanh* smoothing at the phase boundaries. The results are shown as a function of the reduced temperature $r^*$ in figure 3. As expected the interface thickness increases, while the line energy decreases towards the critical point. Considering $r^*$ as a dimensionless temperature, these quantities behave consistently with the expected mean-field critical exponents: we find that for small $|r^*|$, they approach the scaling relationships $d \propto |r^*|^{-0.5}$ and $\gamma \propto |r^*|^{1.5}$, respectively.

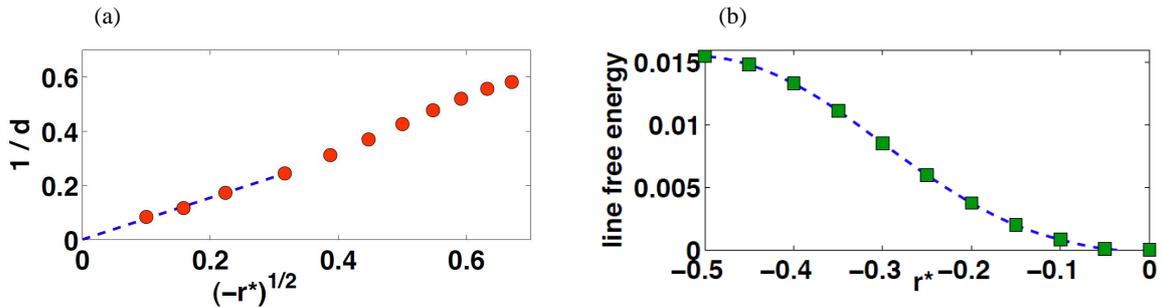

**Figure 3.** Equilibrium interface between the solid and liquid phases in the 2d PFC/SH model. (a) Reciprocal interface thickness vs square root of reduced temperature ($1/d$ vs $|r^*|^{1/2}$); (b) dimensionless line free energy ($\gamma_{SL}$) vs reduced temperature ($r^*$).

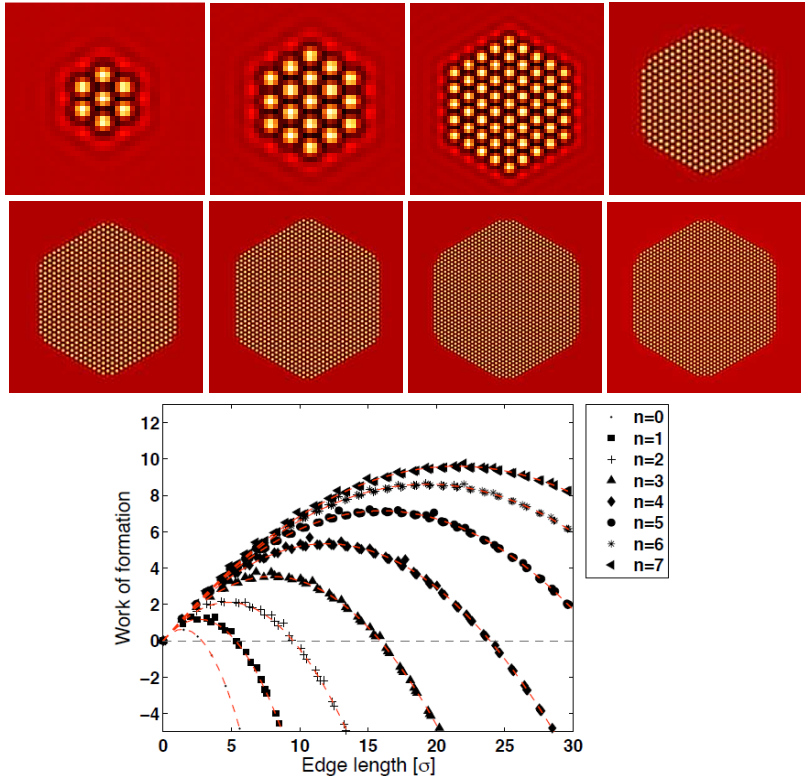

**Figure 4.** Homogeneous nucleation with faceted interfaces in the 2d PFC/SH model at $r^* = -0.5$ and $\psi_0^n = -0.5134 + 0.0134/2^n$, where $n = 0, 1, 2, \ldots, 7$, respectively. Top rows: critical fluctuations (the initial particle density decreases from left to right and from up to down). Bottom panel: nucleation barrier vs size for different initial particle densities.

*3.1.3. Properties of homogeneous nuclei in the 2d PFC/SH model by solving the EL equation.* We have studied nucleation with faceted crystal morphology. To achieve this, our computations have been performed at $r^* = -0.5$, which leads to a strongly faceted interface with excluded orientations (Bäckofen and Voigt 2009, Gránásy *et al* 2010). The initial reduced particle density has been varied so ($\psi_0^n = -0.5134 + 0.0134/2^n$, $n = 0, 1, 2, \ldots, 8$) that the size of nuclei changed substantially. The initial guess for the solution of the EL equation has been constructed as a circle filled with the single-mode solution on a background of homogeneous liquid of particle density $\psi_0^n$ with a *tanh* smoothing at the perimeter. The radius of the circle has been varied in small steps. As opposed to the usual coarse-grained continuum models such as the van der Walls/Cahn-Hilliard/Landau and phase-field type approaches, where the only solutions are the nuclei, here we find a very large number of local extrema of the free energy functional that are all solutions of the EL equation for fixed homogeneous $\psi = \psi_0^n$ in the far field, suggesting that due to the atomistic nature of our clusters the free energy surface is fairly rough.

For small driving forces (large clusters) these solutions appear to map out the nucleation barrier (see figure 4). Since the interface thickness is negligible relative to the cluster size for the larger nuclei, the thermodynamic driving force of crystallization is known, and the shape of the cluster is hexagonal (figure 4), we have applied a version of the classical nucleation theory (see table 1), that assumes a hexagonal shape, to evaluate the line free energy (interface free energy in 2d) from the maximum of the work of formation vs. size relations obtained from a parabolic fit. In analogy to MD results for the hard sphere system (Auer and Frenkel 2001a), the respective effective line free energy increases with increasing driving force (decreasing size). This is attributable to the increasing dominance of the corner energies relative to the line energies for small clusters, whose contribution to the cluster free energy is incorporated into the effective line free energy. Plotting the effective line free energy obtained this way as a function of $1/a$, where $a$ is the length of the sides of the cluster, one observes convergence towards the equilibrium line free energy (figure 5) obtained for a flat boundary in the previous subsection 3.1.2. This suggests that the uncertainties, associated with finding the height and size of the critical fluctuations, are negligible.

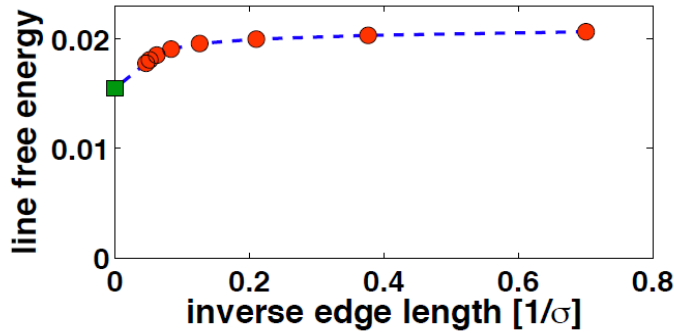

**Figure 5.** Effective line free energy deduced from the work of formation of faceted nuclei in the 2d PFC/SH model at $r^* = -0.5$ as a function of the inverse size (inverse edge length) of nuclei. Note that the data evaluated from the nucleation barrier extrapolate to the value (green square) for the equilibrium (flat) interface.

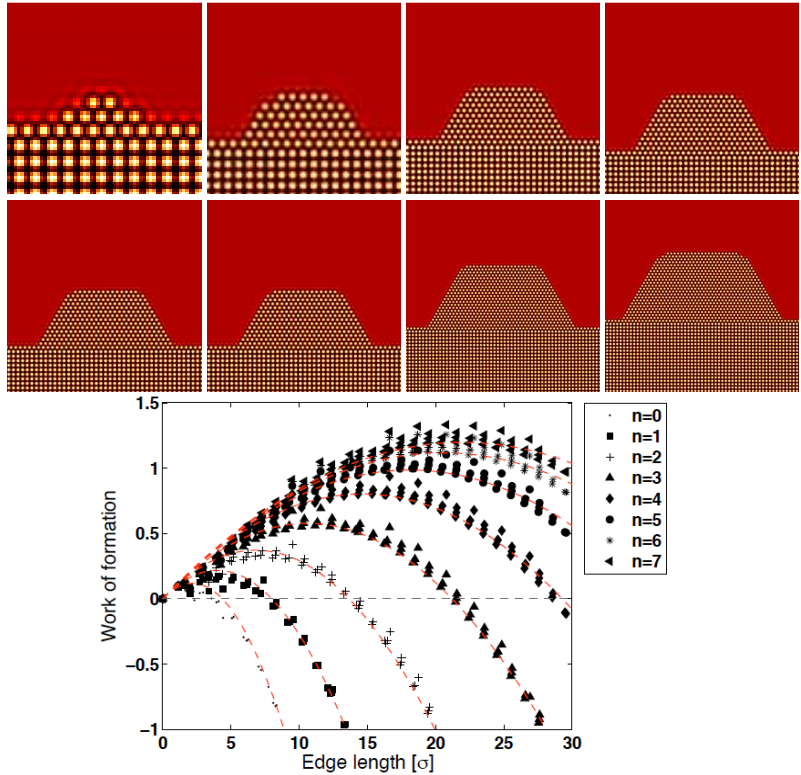

**Figure 6.** Heterogeneous nucleation with faceted interfaces on a square-lattice substrate in the 2d PFC/SH model at $r^* = -0.5$, and ($\psi_0^n = -0.5139 + 0.002/2^n$, where $n = 0, 1, 2, \ldots, 7$, respectively. The lattice constant of the substrate is equal to the interparticle distance in the 2d hexagonal crystal. Top rows: critical fluctuations (the initial particle density decreases from left to right and from up to down). Bottom panel: nucleation barrier vs size for different initial particle densities.

*3.1.4. Properties of heterogeneous nuclei in the 2d PFC/SH model by solving the EL equation.* We have performed a similar analysis for heterogeneous nucleation at the same reduced temperature ($r^* = -0.5$), however, for considerably smaller driving forces ($\psi_0^n = -0.5139 + 0.002/2^n$, $n = 0, 1, 2, \ldots, 8$). The lattice constant of the square-lattice substrate is equal to the interparticle distance of the 2d hexagonal phase. The work of formation of heterogeneous nuclei as a function of size and the image of the crystallites forming at the top of the curves are shown in figure 6. It is remarkable that nuclei are able to form only on the top of a monolayer adsorbed on the surface of the substrate. The formation of such a monolayer substantially decreases the fee energy of the system. The contact angle is 60° determined by the crystal structure, and is apparently decoupled from the substrate by the adsorbed monolayer. Further work is needed to explore how far this observation is true.

*3.1.5. Equilibrium shapes in the 3d single-component PFC model by solving the equation of motion.* Being metastable phases, sufficiently large clusters of the hcp and fcc structure are expected to grow in the absence of noise, just as clusters of the stable bcc phase (Tegze *et al* 2009b). This idea has been used to obtain the equilibrium shape for the bcc, hcp, and fcc crystal structures at the parameter set specified in section 2.4. It has been realized by growing spherical seeds of the required structure until reaching equilibrium with the remaining liquid. The sc crystallite has proven unstable and transformed to bcc fast. We have observed rhombic-dodecahedral, octahedral, and hexagonal-prism shapes for the bcc, fcc, and hcp structures, bound exclusively by the $\{110\}$, the $\{111\}$, and the $\{10\bar{1}0\}$ and $\{0001\}$ faces, respectively (see figure 7). This strong faceting (often seen in colloids: Onoda 1985, Skjeltorp 1987) emerges as a result of a thin crystal-liquid interface that extends to ~ 1 – 2 molecular layers, and has been expected as a result of the large distance form the critical point, leading to a high entropy of transition associated with interface faceting. With the exception of the hcp structure, where $\gamma_{10\bar{1}0}/\gamma_{0001} = 1.08 \pm 0.01$, the specific monoface crystal habits prevent us from evaluat-ing the anisotropy of the interfacial free energy $\gamma_{SL}$ by the Wulff construction. Since the final state of these computations is an equilibrium state, the equilibrium shape obtained this way is also equilibrium shape for the 3d Swift-Hohenberg model.

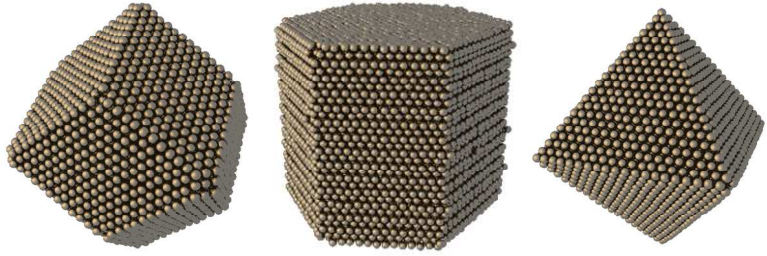

**Figure 7.** Equilibrium shapes the single component PFC model predicts in 3d for the bcc, hcp, and fcc structures, respectively. Spheres of diameter of the interparticle distance, centered at the particle density peaks, are shown. Analogously to 2d (Bäckofen and Voigt 2009, Gránásy *et al* 2010), approaching the critical point, the equilibrium shape converges to a sphere for all three structures. To avoid sticking into metastable states, a small-amplitude noise has been applied. Although these shapes were obtained using the equation of motion, the final state is equilibrium, thus the results apply also to the 3d Swift-Hohenberg model.

*3.1.6. Properties of homogeneous nuclei in the 3d PFC/SH model by solving the EL equation.* We have applied the technique outlined in section 3.1.4 for finding the homogeneous nuclei for the bcc and fcc structures at the parameter set defined in section 2.4. As noted in the previous sub-section, faceted clusters are expected due to the large entropy of transition that applies far from the critical point. We have used different shapes for making an initial guesses for the nuclei, such as sphere, cube, octahedron, and rhombic-dodecahedron. The results obtained for the fcc and bcc structures are presented in figure 8. It appears that if the initial guess for the shape is unfavorable (i.e. it is far from the compact equilibrium shape), the free energy extrema are much higher than for the compact shapes. Therefore, it appears that the spherical and equilibrium shapes provide the best guess for the minima in the free energy surface. Considering the free energy extrema mapped out, it appears that the nucleation barrier is comparable for the bcc and fcc structures. This together with the closeness of the thermodynamic driving forces for the fcc and bcc solidification (Tegze *et al* 2009b) imply that Turnbull's coefficients for the bcc and fcc structures are rather close ($C_{bcc}/C_{fcc} \approx 1$). This finding is in contradiction with recent results for metals from molecular dynamics simulations that predict $C_{bcc}/C_{fcc} \approx 0.53$ (for review see Asta *et al* 2009). We note, however, that the MD results are for low melting entropy materials, whose solid-liquid interface is rough/diffuse on the atomistic scale, as opposed to our high melting entropy case of strongly faceted sharp interface. Faceting is expected in materials of covalent type bonding, where the broken-bond model is a reasonable approximation, an approach that yields comparable Turnbull's coefficients for the bcc and fcc structures (see e.g. Gránásy and Tegze 1991, Gránásy *et al* 1991). Thus our PFC results are consistent with earlier findings for faceted interfaces from the broken-bond model. We expect that for larger $r^*$ values, the PFC results will fall closer to the findings from MD simulations. Work is underway in this direction.

*3.2. Solving the equation of motion(s) in 3d*

In this section, we investigate various dynamic aspects of solidification. Since we apply here conserved dynamics, as opposed to the non-conserved dynamics of the Swift-Hohenberg model, the results presented in this section do not refer to the Swift-Hohenberg model. In the equation of motion of the PFC models, density relaxes diffusively as in colloidal systems. Accordingly, a Mullins-Sekerka type diffusional instability is expected to occur even in the single-component system, whose interaction with crystal anisotropies is expected to lead to the formation of symmetric dendritic structures. Indeed, the formation of dendritic structures has been reported in colloidal systems (Zhu *et al* 1997, He *et al* 1997, Russel *et al* 1997, Cheng *et al* 2002) and has been attributed to the mechanism mentioned above (Russel *et al* 1997). Here we investigate, whether such dendritic structures form in the single-component PFC model. Along this line, first we demonstrate that anisotropic diffusion controlled growth takes place in the PFC, then we attempt to grow dendritic structures. Next, we study whether a precursor phase forms for crystal nucleation in 3d for iron in the framework of the EOF PFC. In agreement with the findings of Berry *et al* (2008a), instantaneous quenching results in the formation of glassy solids. We prepare such glasses and use their structural properties to evaluate an effective pair potential for the PFC model. Finally, we explore solidification in the presence of chemical diffusion.

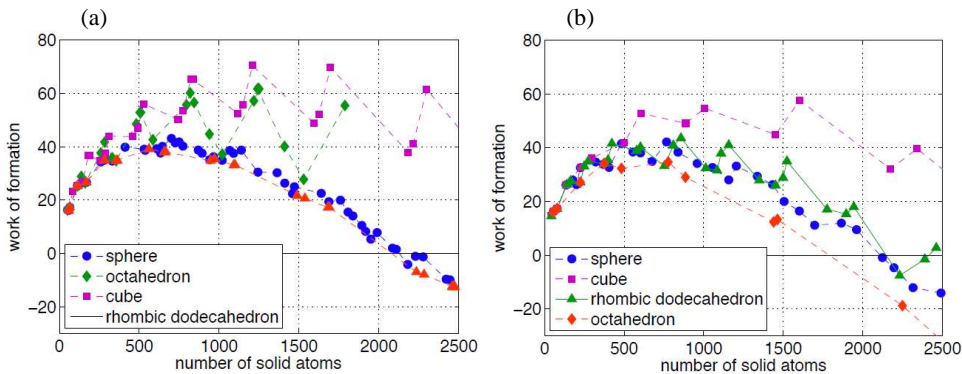

**Figure 8.** Work of formation for the (a) bcc and (b) fcc nuclei as a function of size in the 3d PFC/SH model. Note that the nucleation barriers are comparable, which together with the similarity of the thermodynamic driving forces implies that Turnbull's coefficients for the two phases are comparable.

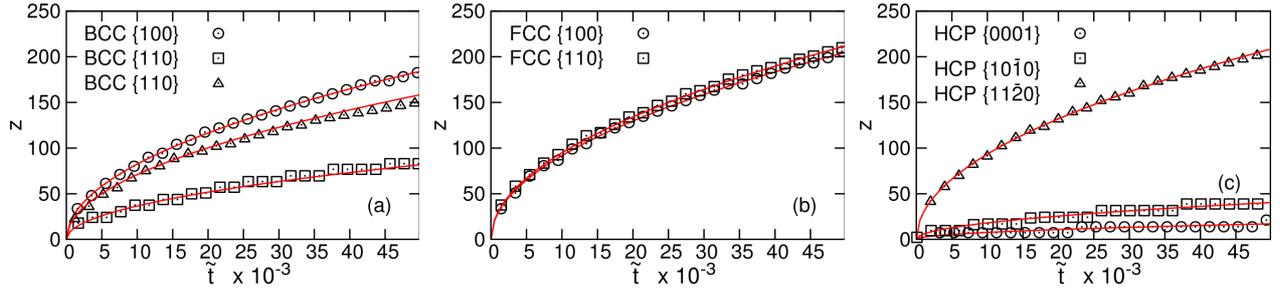

**Figure 9.** Interface position vs dimensionless time for the (a) bcc, (b) fcc and (c) hcp structures in the 3d PFC model obtained with the model parameters by Tegze *et al* (2009b) at $n_0 = -0.04$.

*3.2.1. Diffusion controlled growth in 3d.* Here, we briefly summarize the results we obtained for the growth anisotropy of bcc, hcp, and fcc crystals (Tegze *et al* 2009b). To determine the growth of stress free planar crystal faces, initial crystal slabs have been created so that the linear size in the $x$ and $y$ directions are commensurate with the atomic arrangement of the actual face, while in the $z$ direction the size of simulation box is large enough ($L_z = 1024 \Delta x$) to ensure a period of time, when the diffusion field at the growth front does not yet influence the density at $z = \pm L_z/2$ perceptibly. ($L_x$ and $L_y$ are $\sim L_z/5$.) The position of the front is shown as a function of dimensionless time in figure 9 for the $\{100\}$, $\{110\}$ and $\{111\}$ faces of the bcc and fcc structures, and for the basal $\{0001\}$ and the lateral $\{10\bar{1}0\}$ and $\{11\bar{2}0\}$ faces of the hcp crystal. A closer inspection of the interface region indicates that for these interfaces crystal growth takes place layerwise (Tegze *et al* 2009b). This is reflected in the stepwise change of the position vs. time relationship. After a brief transient, all curves display a roughly $z \propto \tilde{t}^{1/2}$ behavior, indicating a diffusion-controlled growth mechanism, often observed in colloidal systems (e.g. Gast and Monovoukas 1999, Schätzel and Ackerson 1993). To quantify the differences, we have fitted the function $z = z_0 + C(\tilde{t} - \tilde{t}_0)^{1/2}$, to the position − time relationship, where $z_0$ is the initial position, $C$ is the velocity coefficient, and $\tilde{t}_0$ is a transient time. At late times, deviation from this behavior is seen, due to the finite size of the simulation box. Therefore, in the analysis only that part of the growth data have been used, which are free of this effect. The anisotropy of $C$ reflects the differences of the 2D nucleation and step-motion processes on different crystal faces. Such differences have been studied in detail for crystallization from solutions (see e.g. Chernov A A (1989)). We note that the $C$ values presented in table 2 can directly be compared, as they correspond to essentially the same driving force for all the crystalline phases. The bcc, hcp, and fcc sequences for the growth rates are $C_{111} > C_{100} > C_{110}$, $C_{11\text{-}20} > C_{1010} > C_{0001}$ and $C_{110} > C_{100}$. We were unable to determine the growth rate for the fcc $\{111\}$ face, as the hcp $\{0001\}$ interface has started to grow on it because its growth is energetically more favorable. The the hcp $\{0001\}$ interface grows far slower than the other interfaces more corrugated on the atomistic scale. We find that $C$ increases with the driving force differently for the individual faces; i.e., the growth anisotropy varies with supersaturation.

Unfortunately, there appears to be a general lack of experimental data for the anisotropy of diffusion-controlled growth of monatomic bcc, hcp, and fcc crystals in single-component systems. A few examples for the analogous growth of faceted crystals from solutions: The velocity ratio $v_{100}/v_{110} \approx 2.3$ for 3He crystals (bcc) (Tsepelin *et al* 2002), is close to the present $\sim$1.7–2.7, while the ratio $v_{10\_10}/v_{0001\_} \approx 2.8$ observed for Ca(OH)$_2$ (hexagonal but not hcp, Harutyunyan *et*

Table 2. Velocity coefficient $C$ for various interfaces of the bcc, fcc, and hcp structures at $n_0 = -0.04$.

| Structure | {100} | {110} | {111} |
|---|---|---|---|
| bcc | 0.824 ± 0.002 | 0.474 ± 0.005 | 0.948 ± 0.003 |
| fcc | 0.916 ± 0.003 | 0.948 ± 0.002 | - |
| | {10–10} | {11–20} | {0001} |
| hcp | 0.228 ± 0.002 | 0.940 ± 0.002 | 0.096 ± 0.002 |

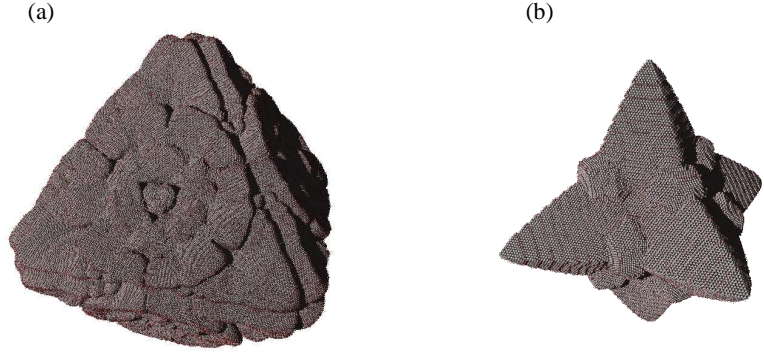

**Figure 10.** Three dimensional dendritic crystals of (a) bcc and (b) fcc structure grown with the model parameters used by Tegze *et al* (2009b). The bcc dendrite has been grown on a 2048 × 2048 × 2048 grid at $n_0 = -0.02$, while the fcc crystal on a 1024 × 1024 × 1024 grid at $n_0 = -0.03$.

*al* 2009) accords reasonably with our ~2.4 for hcp. However, this agreement might be fortuitous. The molecular dynamics simulations indicate a relatively small kinetic anisotropy for the bcc structure, and the sequence of growth velocities varies with the applied potential (Asta *et al* 2009), although usually $v_{100} > v_{110}$ as observed in our PFC study. The MD sequence for the hcp structure obtained for magnesium (Xia *et al* 2007) agrees with our PFC result, however, the anisotropy is smaller. Simulations for the fcc structure (Lennard-Jones system, Ni, Ag, Au, Cu, and Fe) indicate that {100} and {111} orientations have the highest and the lowest growth rates, respectively. The velocity ratio $v_{100}/v_{110}$ varies in the range of 1.2–1.8, as opposed to our PFC result $C_{100}/C_{110} = 0.97$ obtained at $n_0 = -0.04$. These differences are attributable to various reasons: (i) unlike in MD simulations, we have diffusion-controlled growth here; and (ii) the MD simulations refer to materials of low melting entropy ($S_f \sim k_B$), whose crystal-liquid interface extends to 4–5 atomic layers, whereas with the present model parameters the PFC realizes a sharp interface. Increasing the reduced temperature ($r^*$), the PFC predicts more diffuse interfaces.

*3.2.2. Dendritic solidification in 3d.* Large-scale simulations starting with bcc and fcc seeds lead to the formation of dendritic structures. The growth morphologies obtained on a 2048 × 2048 × 2048 grid for the bcc structure at $n_0 = -0.02$ and on a 1024 × 1024 × 1024 grid for the fcc structure at $n_0 = -0.03$ are shown in figure 10. These simulation boxes contain ~ 24 million and ~ 3 million particles, respectively, and correspond to linear sizes of ~ 0.32 mm and ~ 0.16 mm, if $\sigma = 1$ μm is assumed for the diameter of the colloid particles. The bcc dendrite has a rather complex compact octahedral shape with 4-fold split dendrite tips and concentric undulations on the {111} face. The fcc dendrite has a relatively slender, simpler strongly faceted growth morphology. The actual dendrites contain ~ 4.6 and ~ 0.5 million particles, respectively. These sizes are comparable to those of the colloidal dendritic structures grown in microgravity experiments (Zhu *et al* 1997, Russel *et al* 1997, Cheng *et al* 2002). Note that it is the fcc dendrite, whose morphology is close to the shape seen in experiments, which refers to rhcp crystals (a random mixture of fcc and hcp structures).

*3.2.3. Homogeneous nucleation in Fe in 3d.* To generate driving force for solidification at the melting point, we have increased the density/pressure of the Fe liquid until observing nucleation of a solid phase. On the short time scale of our simulations, this has been achieved at extremely high densities: $n_0 \geq 0.5125$, which are both inaccessible experimentally, and are out of the validity range of some of the approximations of the PFC model. Accordingly, the present results need to be taken with precautions.

At $n_0 \geq 0.5125$, an amorphous solid phase nucleates first and grows (indicating a first-order transition), which then transforms first into a polycrystalline bcc phase (and later into a bcc single crystal). This two-step crystallization process is shown in figure 11, which displays the evolution of the atomic configuration and presents structural analysis in terms of the local order parameter $q_6$ that is able to monitor the presence of various crystal structures. (For the definition see e.g. ten Wolde *et al* 1996. Note that for perfect crystals $q_6 = 0.575$ (fcc); 0.485 (hcp); 0.511 (bcc) and 0.354 (sc).) The sequence in figure 11 shows that after an apparently 1st order transition to glass, bcc crystallization takes place. For $n_0 \geq 0.5125$ all these transitions take place in less than 1500 time steps and a polycrystalline state forms. In contrast, we have not detected any phase transition for more than a million time steps at $n_0 = 0.51$. These findings strongly indicate that crystal nucleation is enhanced by the presence of the amorphous precursor, and bcc crystal nucleation directly from the liquid phase requires several orders of magnitude longer time than via the precursor. While we are unaware of experimental evidence for the presence of an amorphous precursor in metallic systems, non-crystalline precursors occur in colloidal systems in 2d (Zhang and Liu 2007, Savage and Dinsmore 2009, DeYoreo 2010) and 3d (Schöpe *et al* 2006, 2007, Iacopini *et al* 2009a, 2009b). We also note in this respect that in an MD study relying on the Ercolessi-Adams embedded atom potential for Al, an amorphous phase has been reported that forms from the liquid by a first-order transition (Mendelev *et al* 2006). Extension of the present nucleation studies for large undercoolings at ambient pressure is underway.

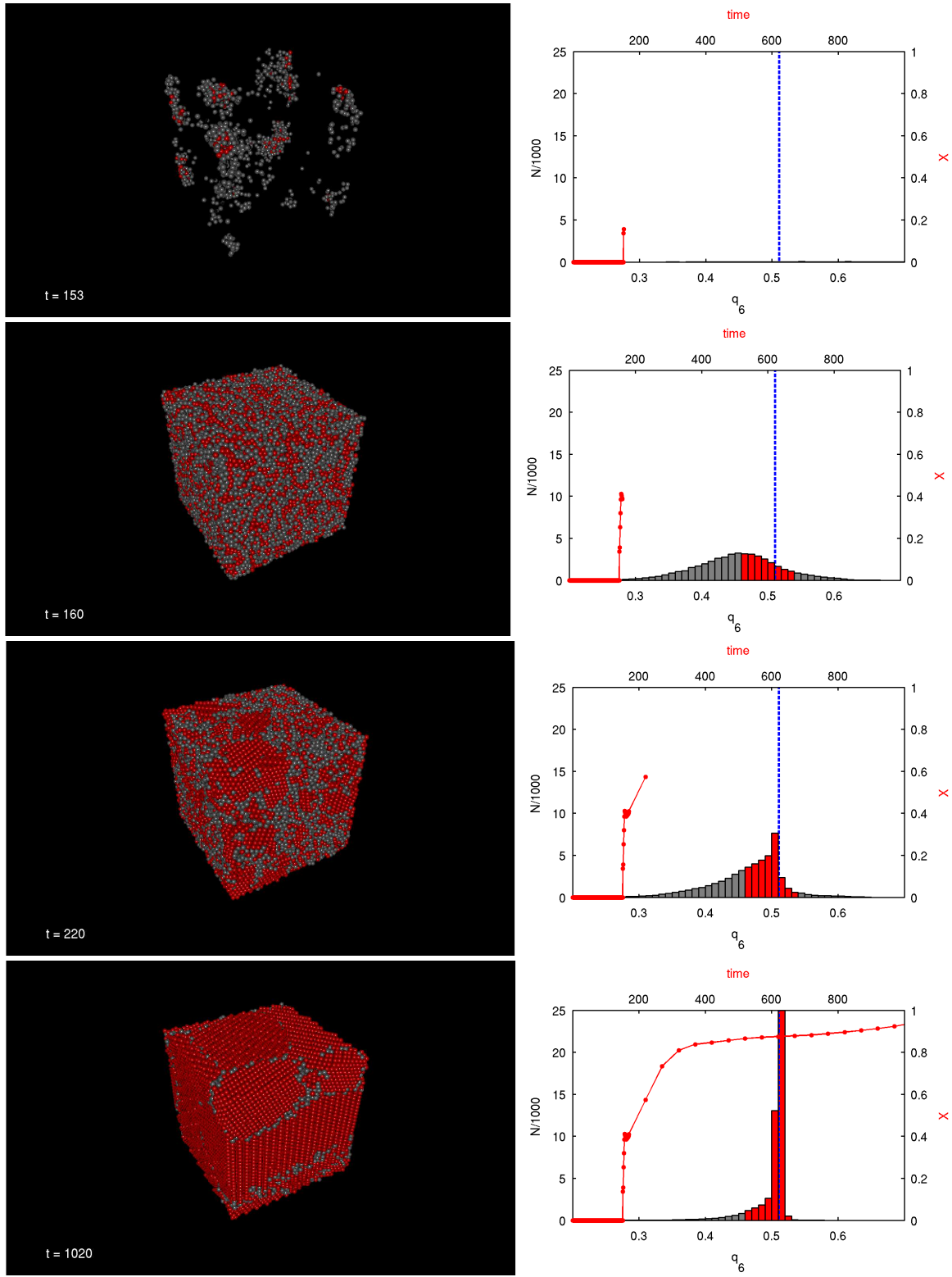

**Figure 11.** Snapshots of two-stage crystallization of highly compressed Fe melt ($n_0 = 0.52$) at the normal pressure melting temperature as predicted by the PFC EOF model (left) and the bcc-like fraction vs time (red) and probability distribution of the structural order parameter $q_6$ (histogram). From top to bottom, the images/graphs correspond to dimensionless times 76.5, 80, 110 and 510. The simulation has been performed on a rectangular grid of size $256 \times 256 \times 256$. Amorphous and bcc surroundings are colored grey and red, respectively. The vertical dashed blue line stands for the value of $q_6$ corresponding to the ideal bcc structure. Note the nucleation of the amorphous phase, its growth until full solidification, and the subsequent crystallization yielding a polycrystalline final state.

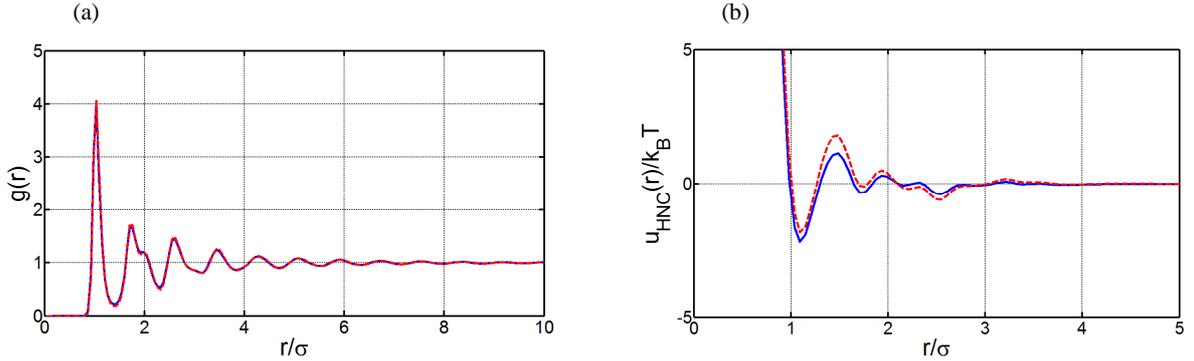

**Figure 12.** Effective pair potential for the 3d PFC model from structural data for the glassy state: (a) radial distribution function $g(r)$ for time steps 1700 (blue) and 10 400 (red dashed) and (b) the respective effective pair potentials derived using the hypernetted chain approximations. Note the complex shape of the pair-potentials and the qualitative resemblance to the DLVO potential often used to model charged colloids.

*3.2.4. Assessment of interparticle potential for the PFC model from the structural properties of glass.* One of the intriguing questions regarding the PFC model is the type of molecular interaction it indeed realizes. Since the physical information that enters the theory is a direct correlation function Taylor expanded in the Fourier space, which diverges for $k \to \infty$, it is not immediately straightforward, what kind of interaction it does impose between the particles. In the present work, we attempt to use the structural properties of the glassy state to deduce an effective pair potential for the PFC model. This is motivated by the fact that pair potentials have been evaluated for simple liquids (such as metals (Simoji 1977) and colloids (Havemann et al 1995, Fritz-Popovski 2009) from structural data using the Percus-Yevick (PY, 1958) and/or hypernetted chain (HNC, van Leeuwen et al 1959) approximations (see e.g. Shimoji 1977). Starting from the Ornstein-Zernike (1914) equation and the PY and HNC closures, the interparticle potential can be expressed in terms of the direct correlation function $c(r)$ and the pair correlation function $g(r)$ as

$$u_{PY}(r)/k_B T = \log\{1 - c(r)/g(r)\} \tag{24}$$

and

$$u_{HNC}(r)/k_B T = g(r) - c(r) - 1 - \log\{g(r)\}. \tag{25}$$

While the PY closure is more appropriate for short range interactions, the HNC is preferred in the case of long range interactions. We have used $1024 \times 1024 \times 1024$ size simulations (~ 3 million particles) to evaluate $g(r)$ in the glassy state (see figure 12a) that forms after instantaneous quenching to the dimensionless temperature $\Delta B = 0.005$ at the initial density $n_0 = 0$ with the parameter set used by Berry et al (2008a). Its Fourier transform has been used to obtain the structure factor $S(k)$, and $c(r) = 1 - 1/S(k)$, which has been then Fourier transformed to compute the real-space direct correlation function $c(r)$. The HNC pair potentials corresponding to 1700 and 10 400 time steps are shown in figure 12b, displaying minor relaxation effects on the effective pair potential. (The PY estimate is in a qualitative agreement with the HNC, however, with spurious cusps, whose position is sensitive to fine details of the evaluation process.) The HNC pair potentials are fairly complex and show several minima resembling to MD simulation results for colloids (Havemann et al 1995). They display some qualitative resemblance to the DVLO potential in the sense that besides the main attractive part there are weak outer minima.

*3.2.5. Eutectic solidification in 2d and 3d.* The ability of the PFC model to describe binary eutectic solidification in 2d has been demonstrated recently (Elder et al 2007, Tegze et al 2009a). In conventional isothermal phase-field theoretical (PFT) simulations, which neglect density difference, eutectic colonies have been seen to form in only systems consisting of three (or higher number of) components (Plapp and Karma 2002). In such cases, the formation of the colonies is associated with morphological instability due to the long-range diffusion field of the third component at the interface, which is evidently absent in the binary case, where only the short-range diffusion mode, parallel with the interface, occurs. Contrary to this, we have observed eutectic colony formation in the binary PFC model (see figure 13). A clue to understand this seemingly counterintuitive finding is given by the observation that, in our simulations, after an initial period of constant velocity, the growth velocity continuously decreases due to the formation of a depletion zone in the total particle density $n$ ahead of the growth front (particle density is larger in the solid). Thus, the propagation of the eutectic front is controlled here by long-range diffusion; a finding that follows from the fact that (at least for small driving forces) the relaxation of $n$ is controlled by particle diffusion in the PFC model. To make the analogy with the conventional phase-field theory of ternary solidification, we note that in the ternary case the PFT consists of three independent fields: a non-conserved field (the structural order parameter or phase-field), plus two conserved fields (the two independent concentrations). As opposed to this, in the binary case (where no colony formation has been observed), the PFT consist of a single non-conserved field that is coupled to a conserved one. The PFC model, however, considers the density change during freezing, and this change of the local density happens via diffusion. Accordingly, the situation described by a binary PFC model can be represented by three coupled fields in the language of conventional PFT: a non-conserved structural order parameter, and two conserved fields: the concentration field, and the total particle density field. As a result, the conditions realized by the PFC model are mathematically analogous to those of the usual ternary PFT, thus one indeed expects the formation of eutectic colonies. We note that this mode of binary eutectic colony formation is expected to occur only in colloidal systems, where density relaxation is indeed diffusive. Unfortunately, experimental realization of eutectic solidification in colloids is far from being trivial (Lorenz *et al* 2008, 2009a, 2009b).

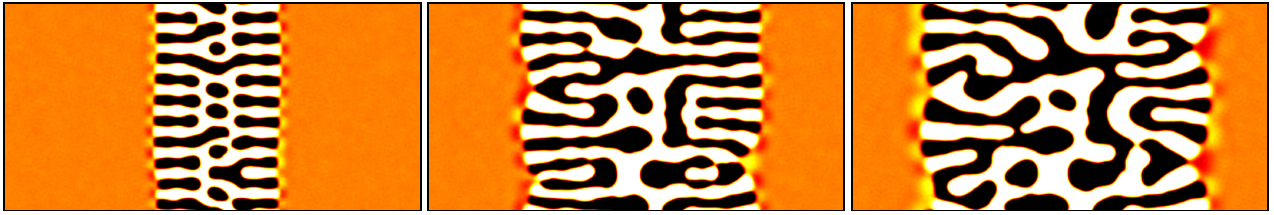

**Figure 13.** Snapshots of eutectic solidification on the atomistic scale in the binary PFC model in 2d: Composition ($\delta N$) maps corresponding to $2 \times 10^5$, $6 \times 10^5$ and $10^6$ time steps are shown. White and black denote the two crystalline phases, while yellow stands for the liquid phase. The simulation has been performed on a $2048 \times 1024$ rectangular grid. Crystallization has been started by placing a row of supercritical crystalline clusters of alternating composition into the simulation window. Interestingly, the eutectic pattern evolves inside the solid region on a timescale comparable to the timescale of solidification.

Finally, we have performed illustrative simulations in 3d for eutectic solidification that has been started by placing a two-phase seed into the simulation box composed of the two coexisting bcc phases. A sequence of snapshots, showing the time evolution of solidification, is displayed in figure 14. Remarkably, at the large driving force realized by the applied conditions, growth takes place at a high velocity that leads to freezing with a non-equilibrium density. Details of this "density trapping" process, which is analogous to solute trapping observed during rapid solidification of alloys (see e.g. Aziz 1982 and Jackson *et al* 2004), are discussed elsewhere (Gránásy *et al* 2010, Tegze *et al* 2010).

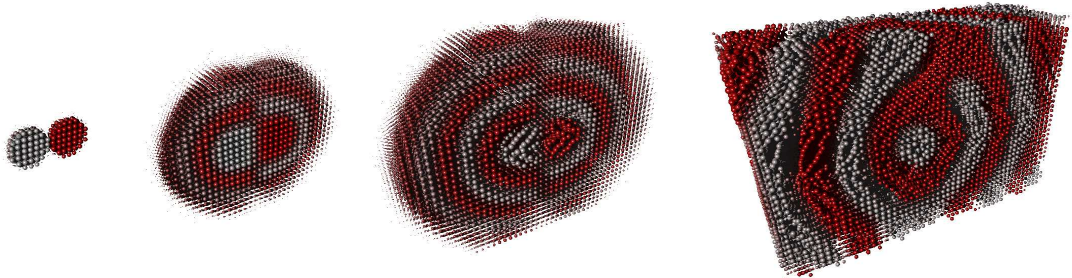

**Figure 14.** Snapshots of eutectic solidification on the atomistic scale as predicted by the binary PFC model in 3d: Time elapses from left to right. The simulation has been performed on a $450 \times 300 \times 300$ rectangular grid. The simulation has been started by placing two touching supercritical bcc clusters of different compositions into the simulation window. Note the continuous bcc structure forming initially, which breaks up to lamellae separated by lower density phase boundaries at later stages of the simulation. Remarkably, the nanoscale solid-phase eutectic pattern roughens on a timescale comparable to the time of solidification. The brown and grey colors denote the terminal solutions of the two crystalline phases. Spheres of size reflecting the height of the local total number density peak ($n$) and colored according to the local composition ($\delta N$) are centered to the particle density maxima. Only half of the simulation window is shown ($450 \times 150 \times 300$).

## 4. Summary

We have used the phase-field crystal (PFC) method to explore polymorphism and various aspects of crystal nucleation and growth in two and three dimensions. More specifically, in the present paper we have

- refined the 3d phase diagram of the one-component PFC/Swift-Hohenberg model,
- determined the equilibrium interfacial properties in the 2d PFC/Swift-Hohenberg model,
- evaluated the nucleation barrier for homogeneous and heterogeneous crystal nucleation in the 2d and 3d PFC/Swift-Hohenberg models,
- explored the anisotropy of growth rate for diffusion-controlled layerwise solidification,
- shown that, due to the diffusional dynamics of density relaxation the PFC model assumes, dendrites can be grown in isothermal single-component systems,

- demonstrated that according to the EOF PFC model crystal nucleation in compressed Fe liquid happens via an amorphous precursor,
- evaluated an effective pair potential for PFC from the 3d glass structure using the Percus-Yevick and hypernetted chain approximations,
- performed illustrative simulations for eutectic solidification and shown that due to the diffusive dynamics the PFC model assumes for the total number density, eutectic colonies form in our binary systems.

These results imply that the PFC model is a flexible tool for studying the microscopic aspects of crystalline freezing. Work is underway to extend the present studies for more complex cases of the substrate-crystal interaction and crystal nucleation, and for further exploration of the model's applicability to real materials.

## Acknowledgments

We thank Akusti Jaatinen and Tapio Ala-Nissila (Helsinki University of Technology, Espoo, Finland) for the valuable discussions on the thermodynamics of the single component PFC systems, and especially for sending us their manuscripts on the 3d phase diagram and on the EOF PFC model prior to publication. We thank Mathis Plapp (École Polytechnique, CNRS, Palaiseau, France) for the enlightening discussions on noise and Miklós Tegze (Research Institute for Sold State Physics and Optics, Budapest, Hungary) for his expert advices concerning the evaluation of the structure factor for the amorphous phase. This work has been supported by the EU FP7 Collaborative Project ENSEMBLE under Grant Agreement NMP4-SL-2008-213669 and by the Hungarian Academy of Sciences under contract OTKA-K-62588. TP is a grantee of the Bolyai János Scholarship.

## References

Achim C V, Karttunen M, Elder K R, Granato E, Ala-Nissila T and Ying S C 2006 *Phys. Rev*. E **74** art no 021104
Alexander S and McTague J 1978 *Phys. Rev. Lett.* **41** 702
Archer A and Rauscher M 2004 *J. Phys. A* **37** 9325
Asta M, Beckermann C, Karma A, Kurz W, Napolitano R, Plapp M, Purdy G, Rappaz M and Trivedi R 2009 *Acta Mater.* **57** 941
Auer S and Frenkel D 2001a *Nature* **409** 1020
— 2001b *Nature* **413** 71
— 2003 *Phys. Rev. Lett.* **91** art no 015703
Aziz M J 1982 *J. Appl. Phys.* **53** 1158
Bäckofen R and Voigt A 2009 *J. Phys.: Condens. Matter.* **21** art no 464109
Bartell L S and Wu D T 2007 *J. Chem. Phys.* **127** art no 174507
Berry J, Elder K R and Grant M 2008a *Phys. Rev. E* **77** art no 061506
— 2008b *Phys. Rev. B* **77** art no 224114
Brazovskii S A 1975 *Zh. Eksp. Teor. Fiz.* **68** 175
Cheng Z D, Chaikin P M, Zhu J X, Russel W B and Meyer W V 2002 *Phys. Rev. Lett.* **88** art no 015501
Chernov A A 1989 *Contemp. Phys.* **30** 251
Christian J W 1981 *Transformations in Metals and Alloys.* (Oxford: Pergamon)
DeYoreo J J 2010 to be published
Elder K R, Katakowski M, Haataja M and Grant M 2002 *Phys. Rev. Lett.* **88** art no 245701
Elder K R and Grant M 2004 *Phys. Rev. E* **70**, art no 051605
Elder K R, Provatas N, Berry J, Stefanovic P and Grant M 2007 *Phys. Rev. B* **75**, art no 064107
Elenius M and Dzugutov M 2009 *J. Chem. Phys.* **131** art no 104502
Esztermann A and Löwen H 2005 *J. Phys.: Condens. Matter* **17** S429
Frigo M and Johnson S G 2005 *Proc. IEEE* **93** 216
Fritz-Popovski G 2009 *J. Chem. Phys.* **131** art no. 114902
Gast A P and Monovoukas Y 1991 *Nature* **351** 553
Greer A L, Bunn A M, Tronche A, Evans P V and Bristow D J 2000 *Acta Mater.* **48** 2823
Gránásy L and Tegze M 1991 *Mater. Sci. Forum* **77** 243
Gránásy L, Tegze M, and A Ludwig 1991 *Mater. Sci. Eng. A* **133** 577
Gránásy L, Tegze G, Tóth G I and Pusztai T 2010 submitted to *Philos. Mag.*, arXiv:1003.0454v1 [cond-mat.mtrl-sci]
Gross N A, Ignatiev M and Chakraborty B 2000 *Phys. Rev. E* **62** 6116
Haataja M, Gránásy L and Löwen H 2010 this volume
Harutyunyan V S, Kirchheim A P, Monteiro P J M, Aivazyan A P and Fischer P 2009 *J Mater. Sci.* **44** 962
Havemann U, Grivtsov and Merkulenko N N 1995 *J. Chem. Phys.* **99** 15518
He Y, Olivier B and Ackerson B J 1997 *Langmuir* **13** 1408
Herlach D M 1994 *Mater. Sci. Eng. Rev.* **12** 177
Herlach D M, Holland-Moritz D and Galenko P 2007 *Metastable Solids from Undercooled Melts* (Amsterdam: Elsevier)
Herring C 1951 *The Physics of Powder Metallurgy*, ed W. E. Kingston, (New York: McGrew-Hill) Chap. 8, p. 143.
Iacopini S, Palberg and Schöpe H J 2009a *J. Chem. Phys.* **130** art no 0845502
— 2009b *Phys. Rev. E* **79** art no 010601(R)
Jaatinen A and Ala-Nissila T 2010 this volume
Jaatinen A, Achim C V, Elder K R and Ala-Nissila T 2009 *Phys. Rev. E* **80** art no 031602

Jackson K A, Beatty K M and Gudgel K A 2004 *J. Cryst. Growth* **271** 481
Johnson T A and Elbaum C 2000 *Phys. Rev. E* **62** 975
Karma A 2009 personal communication
Karma A and Rappel W-J 1999 *Phys. Rev. E* **60** 3614
Klein W 2001, *Phys. Rev. E* **64** art no 056110
Lorenz N J, Liu J and Palberg T 2008 *Colloid and Surfaces* **319** 109
Lorenz N J, Schöpe H J, Reiber H, Palberg T, Wette P, Klassen I, Holland-Moritz D, Herlach D M and Okubo T 2009a *J. Phys.: Condens. Matter* **21** art no 464116
Lorenz N J, Schöpe H J and Palberg T 2009b *J. Chem. Phys.* **131** art no 134501
Löwen H 2003 *J. Phys.: Condens. Matter* **15** V1
Lutsko J F and Nicolis G 2006 *Phys. Rev. Lett.* **96** art no 046102
Majaniemi S 2009 personal communication
Majaniemi S and Provatas N 2009 *Phys. Rev. E* **79** art no 011607
Marconi U M B and Tarazona P 1999 *J. Chem. Phys.* **110** 8032
Mendelev M I, Schmalian, Wang C Z, Morris J R and Ho K M 2006 *Phys. Rev. B* **74** art no 104206
Mellenthin J, Karma A and Plapp M 2008 *Phys. Rev. B* **78** art no 184110
Onoda G Y 1985 *Phys. Rev. Lett.* **55** 226
Ornstein L S and Zernike F 1914 *Verh.-K. Ned. Akad. Wet., Afd. Natuurkd.,* Eerste Reeks **17** 793
Percus J K and Yevick, G J 1958 *Phys. Rev.* **110** 1
Plapp M 2010 submitted to *Philos. Mag.*
Plapp M and Karma A 2002 *Phys. Rev. E* **66** art no 061608
Prieler R, Hubert J, Li D, Verleye B, Haberkern R and Emmerich H 2009 *J. Phys.: Condens. Matter* **21** art no 464110
Provatas N, Dantzig J A, Athreya B, Chan P, Stefanovic P., Goldenfeld N and Elder K R 2007 *JOM* **59** 83
Pusztai T, Tegze G, Tóth G I, Környei L, Bansel G, Fan Z and Gránásy L 2008 *J. Phys.: Condens. Matter* **20** art no 404205
Quested T D and Greer A L 2005 *Acta Mater.* **53** 2683
Ramakrishnan T V and Yussouff M 1979 *Phys. Rev. B* **19** 2775
Reavley S A and Greer A L 2008 *Philos. Mag.* **88** 561
Russel W B, Chaikin P M, Zhu J, Meyer W V and Rogers R 1997 *Langmuir* **13** 3871
Savage J R and Dinsmore A D 2009 *Phys. Rev. Lett.* **102** art no 198302
Schätzel K and Ackerson B J 1993 *Phys. Rev. E* **48** 3766
Schöpe H J, Bryant G and van Megen W 2006 *Phys. Rev. Lett.* **96** art no 175701
——     2007 *J. Chem. Phys.* **127** art no 084505
Sear R P 2001 *J. Chem. Phys.* **114** 3170
Shimoji M 1977 *Liquid Metals* (London: Academic Press) Chaps. 105 and 304
Shiryayev A and Gunton J D 2004 *J. Chem. Phys.* **120** 8318
Skjeltorp A T 1987 *Phys. Rev. Lett.* **58** 1444
Skripov V P 1976 *Crystal Growth and Materials*, ed E Kaldis and H J Scheel (Amsterdam: North Holland) p 327
Swift J and Hohenberg P C 1977 *Phys. Rev. A* **15** 319
Swope W C and Andersen H C 1990 *Phys. Rev. B* **41**, 7024
Talanquer V and Oxtoby D W 1998 *J. Chem. Phys.* **109** 223
Tegze G, Bansel G, Tóth G I, Pusztai T, Fan Z and Gránásy L 2009a *J. Comput. Phys.* **228** 1612
Tegze G, Gránásy L, Tóth G I, Podmaniczky F, Jaatinen A, Ala-Nissila T and Pusztai T 2009b *Phys. Rev. Lett.* **103** art no 035702
Tegze G, Gránásy L, Tóth G I, Douglas J. F. and Pusztai T 2010 to be published
ten Wolde P R and Frenkel D 1997 *Science* **277** 1975
——     1999 *Phys. Chem. Chem. Phys.* **1** 2191
ten Wolde P R, Ruiz-Montero M J and Frenkel D 1995 *Phys. Rev. Lett.* **75** 2714
——     1996 *J. Chem. Phys.* **104** 9932
Tóth G I and Gránásy L 2007 *J. Chem. Phys.* **127** 074710
Tóth G I and Tegze G 2010 to be published
Toxwaerd S 2002 *J. Chem. Phys.* **117** 10303
Tsepelin V, Alles H, Babkin A, Jochemsen R, Parshin A. Y. and Todoshchenko 2002 *Phys. Rev. Lett* **88** art no 045302
van Leeuwen J M J, Groeneveld J and de Boer J 1959 *Physica (Amsterdam)* **25** 792
van Teeffelen S, Likos C N and Löwen H 2008 *Phys. Rev. Lett.* **100** art no 108302
van Teeffelen S, Backofen R, Voigt A and Löwen H 2009 *Phys. Rev. B* **79** art no 051404
Wang H, Gould H and Klein W 2007 *Phys. Rev. E* art no 031604
Webb III E B, Grest G S and Heine D R 2003 *Phys. Rev. Lett.* **91** art no 236102
Wu K A and Karma A 2007 *Phys. Rev. B* **76** art no 184107
Woodhead-Galloway J and Gaskell T 1968 *J. Phys. C* **1** 1472
Xia Z G, Sun D Y, Asta M and Hoyt J J 2007 *Phys. Rev. B* **75** art no 012103
Yasuoka K, Gao G T, and Zeng X C 2000 *J. Chem. Phys.* **112** 4279
Zhang T J and Liu X Y 2007 *J. Am. Chem. Soc.* **129** 13520
Zhu J, Li M, Rogers R, Meyer W, Ottewill R H, STS-73 Space Shuttle Crew, Russel W B and Chaikin P M 1997 *Nature* **387** 883